\documentclass[prb,showpacs]{revtex4}
\usepackage{amsmath}
\usepackage{amssymb}
\usepackage[dvips]{graphicx}
\newcommand{\be}{\begin{equation}}
\newcommand{\ee}{\end{equation}}
\newcommand{\bfig}{\begin{figure}[!ht]\begin{center}}
\newcommand{\efig}{\end{center}\end{figure}}
\newcommand{\ba}{\begin{eqnarray}}
\newcommand{\ea}{\end{eqnarray}}
\newcommand{\ncs}{\sum\limits_{i=1}^{L}\sum\limits_{j=1}^{L}}

\begin{document}
\newcommand*{\acro}[1]{{\small\textsc{#1}}}
\newcommand*{\lit}[1]{\texttt{#1}}
\newcommand*{\pkg}[1]{\textsf{#1}}

\title{Dynamics of Desorption with Lateral Diffusion}

\author{Tjipto Juwono$^{a,b}$}
\author{Per Arne Rikvold$^a$ (E-mail: prikvold@fsu.edu)}
\affiliation{$^a$ Department of Physics, Florida State University, 
Tallahassee, Florida 32306-4350, USA\\
$^b$ Surya University, Tangerang 15810, Banten, Indonesia}
\date{\today}

\begin{abstract}
The dynamics of desorption from a submonolayer of adsorbed
atoms or ions are significantly
influenced by the absence or presence of lateral diffusion of the
adsorbed particles. When diffusion is present, the adsorbate configuration
is simultaneously changed by two distinct processes, proceeding in parallel:
adsorption/desorption, which changes the total adsorbate coverage, and
lateral diffusion, which is coverage conserving. Inspired by experimental
results, we here study the effects of these competing processes by kinetic
Monte Carlo simulations of a simple lattice-gas model.
In order to untangle the various effects, we perform large-scale simulations,
in which we monitor coverage, correlation length, and
cluster-size distributions, as well as the behavior of representative individual clusters, 
 during desorption. For each initial
adsorbate configuration, we perform multiple, independent simulations, 
without and with diffusion, respectively.
We find that, compared to desorption without diffusion, the coverage-conserving
diffusion process produces two competing effects: 
a retardation of the desorption rate, which is associated with a 
coarsening of the adsorbate configuration, 
and an acceleration due to desorption of monomers ``evaporated" 
from the cluster perimeters. 
The balance between these two effects is
governed by the structure of the adsorbate layer at the beginning of the
desorption process. Deceleration and coarsening are predominant for 
configurations dominated by monomers and small clusters, 
while acceleration is predominant for
configurations dominated by large clusters. 
\end{abstract}
\maketitle

\section{Introduction}
\label{sec-intro}

Island growth and dissolution on surfaces 
are important non-equilibrium problems, both from 
fundamental and technological points of view.  The interplay of
adsorption, desorption, and lateral 
diffusion of adsorbate particles is essential to understanding cluster 
dynamics, and it has therefore been extensively 
studied.\cite{HON67,ABR70,GIL72,GIL76,BAR92,BAR93,BAR95,BAR96,MOR96,MOR99,LEH00,LEH01,HEY01,YAO02,WAT04,FRA05,FRA06,KOD06,ROO08,NGU08,JUW13}
The present investigation is inspired by two experimental studies.

He and Borguet studied gold cluster formation 
and dissolution on Au(111) electrodes.\cite{HEY01}
In these experiments, during a short positive potential pulse, the 
surface reconstructed and gold atoms 
were released onto the reconstructed surface, where they quickly 
nucleated and formed monolayer 
clusters. After the pulse, the reconstruction was lifted and
gold atoms were reabsorbed  in 
such a way that small clusters tended to decay quickly, while large clusters  
initially continued to  grow before they also eventually decayed.
Island stability in this experiment was studied by monitoring the cluster 
dissolution dynamics.  The overall dissolution dynamics of the  clusters 
was described by plotting the cluster coverage, i.e., 
the fraction of the surface covered by clusters, as a function of time.

Bartelt~{\it et al.}\cite{BAR95,BAR96} made cluster-by-cluster \emph{in situ}
observations of the coarsening process by Ostwald ripening~\cite{LIF61} 
in Si/Si(001) submonolayer systems.
They wanted to understand how each cluster behaves in response to its surroundings.
In this case, atoms detach from clusters, 
diffuse through the two-dimensional adsorbate gas surrounding the clusters,
and eventually attach to other clusters, 
with net flow from smaller to larger clusters.
In general we can expect that the behavior of each cluster depends on 
the detailed configuration of the surrounding clusters.~\cite{BAR96}

Another typical application is the experimental study of underpotential 
deposition (UPD), e.g.,  of Cu on electrodes of Au(111).\cite{ATA00}
In UPD, a submonolayer of one metal is electrochemically adsorbed onto another
in a range of electrode potentials more positive than that 
in which bulk deposition occurs.\cite{RIK98}

A large body of work exists on the effects of diffusion on cluster growth and 
pattern formation in systems undergoing net {\it ad\/}sorption,\cite{MEAK98}
and for systems undergoing coarsening by lateral diffusion at constant 
coverage.\cite{LIF61,KHAR95,KHAR98,MILL99,KARA09}  
Much less is known for systems undergoing net 
{\it de\/}sorption.\cite{KRUG05,FRA05,FRA06,IVAN10} 
In the present study we therefore investigate in a simple lattice-gas model 
the changes in cluster dynamics and cluster size distribution 
that occur during desorption in the presence of lateral  
diffusion as the initial cluster configuration is varied. 
The change in initial configuration would result in the change of the  
surrounding configurations of any particular cluster at a given 
coverage or time. In general, the behavior of each cluster depends on  
the detailed configuration of the surrounding clusters.\cite{BAR96}
The simultaneous action of the nonconserved-order-parameter processes of
adsorption and desorption and the conserved-order-parameter process of lateral
 adparticle diffusion\cite{LIF61} leads to a complex interplay
 between  acceleration and deceleration of the overall desorption process, 
depending on the details of the local
configurations.~\cite{BAR92,BAR96,FRA05,FRA06}

Using Kinetic Monte Carlo (KMC) simulation of a single-layer lattice-gas model, 
we consider the simultaneous adsorption, desorption, and lateral diffusion 
processes. Specifically, we employ the $n$-fold way rejection-free KMC 
algorithm,~\cite{BOR75, NOV01, GIL72, GIL76} 
in which the simulation clock is updated after every accepted move. 
We consider two kinds of elementary moves: (1) an adsorption/desorption 
move, and (2) a single-atom diffusion move to a nearest-neighbor site. 
During the simulation, we sample the time development
 of the adsorbate  coverage  and correlation length. 
At certain coverages or times we also analyze the
clusters using the Hoshen-Kopelman algorithm.~\cite{HOS76} With this
algorithm we are able to measure cluster size distributions and
to tag specific clusters and follow their individual dynamics. 
The restriction to submonolayer configurations is consistent with the 
situation in the experiments reported in Refs.~\onlinecite{HEY01,BAR95,BAR96}.
Some preliminary results of our study 
were previously presented in Ref.~\onlinecite{JUW13}.

The rest of this paper is organized as follows. The model, algorithm, and
the measured quantities are introduced in Sec.~\ref{model}.
We present the procedures to prepare the initial configurations
in Sec.~\ref{preparation}. The results
 are presented in Sec.~\ref{results}, and our conclusions are summarized  
  in Sec.~\ref{analysis}. 

\section{Model, Algorithm, and Measurements}\label{model}

\subsection{Lattice-gas Model}
The simplest kind of model for adsorption, desorption, and lateral diffusion 
on a surface is a lattice-gas model.
We limit our study to submonolayer systems, so that each lattice site 
can be occupied by at most one atom.  In this model, 
we define the occupancy number $c_{i} = 1$ when site $i$ is occupied, 
and $c_{i} = 0$ otherwise.  
Adsorption is the process of  occupying an empty lattice site
by a single particle, while desorption is the opposite. Diffusion 
consists in the random hopping of a single particle to an empty 
nearest-neighbor site. 
We work with square lattices with a total number of
sites  $V=L \times L$ and periodic boundary conditions.
The total number of adsorbed particles is 
$N_p=\sum_{i=1}^{V} c_i$,
and the coverage (or particle concentration) is defined as
\begin{eqnarray}
\theta={N_p}/{V} \;.
\label{eqncove}
\end{eqnarray}

\subsection{Hamiltonian and Kinetic Monte Carlo Algorithm}

The grand-canonical effective Hamiltonian of the lattice-gas model for 
our system is
\begin{eqnarray}
{\mathcal H}=-\phi\sum_{\langle i,j \rangle}c_i(t)c_j(t)-\mu\sum_i c_i(t) \;,
\label{hamil}
\end{eqnarray}
\noindent where $\phi > 0 $ is an attractive nearest-neighbor 
interaction constant, and $\mu$ is the electrochemical potential.  
Below the model's critical temperature $T_c$, the value of the latter at 
coexistence between the 
low-coverage and high-coverage phases is $\mu_0=-2 \phi$. 
The first sum in Eq.~(\ref{hamil}) runs over all nearest-neighbor pairs of 
sites, and the second sum runs over all sites. 
This lattice-gas Hamiltonian is equivalent to a nearest-neighbor 
Ising spin model,~\cite{ISING}
and the adsorption, desorption, and nearest-neighbor diffusion
processes can conveniently be discussed in Ising language,  
using single spin-flip and Kawasaki nearest-neighbor spin-exchange moves.

The simulations were performed for  a $256 \times 256$ square 
lattice.  The temperature $T$ equals $0.8T_c$
with $T_c$ the exact critical temperature of the Ising  
lattice-gas model.~\cite{ONS44}
This temperature is low enough to avoid complications 
due to critical fluctuations, while it is high enough to 
obtain multidroplet configurations.~\cite{RIK94,FRA06}
The electrochemical potential and temperature are 
hereafter given in dimensionless 
units of the interaction constant $\phi$. (Boltzmann's constant $k_B=1$.)
The length unit is the lattice constant of the two-dimensional simulation 
lattice. 

To obtain simulations that are closer to a real physical system, we
consider the fact that
when the system goes from a state of energy $E_{\rm A}$ to $E_{\rm B}$, it
has to overcome an energy barrier (see Fig.~\ref{picbarr}). 
We define the height
of the energy barrier ${\Delta}$ of the symmetric Butler-Volmer
type\cite{FRA06} as
\begin{eqnarray}
\Delta 
       &=& E_{\rm H} - \frac{1}{2}(E_{\rm A} + E_{\rm B}) \;,
\label{eqndelt}
\end{eqnarray}
where $E_{\rm H}$ is the maximum energy (saddle-point energy) along 
the transition path. 
The energy difference between $E_{\rm H}$ and $E_{\rm A}$ is therefore 
\begin{eqnarray}
\Delta {\tilde E} &=& E_{\rm H} - E_{\rm A} \nonumber \\
                  &=& \frac{1}{2}(E_{\rm B} - E_{\rm A}) + \Delta.
\label{eqndele}
\end{eqnarray}
Different processes have different barrier values: 
adsorption/desorption has barrier $\Delta_{\rm ads/des}$, and lateral 
diffusion has $\Delta_{\rm d}$.
Since our objective is to study the competition between desorption and 
diffusion, for simplicity we use the same barrier $\Delta_{\rm d}$ 
for all diffusion moves, regardless of
whether they occur along or away from a cluster 
edge.\cite{KHAR95,KHAR98,MILL99,YILD09}  
To study the effects of the diffusion on the desorption dynamics, 
we vary the difference $\left(\Delta_{\rm d} - \Delta_{\rm ads/des} \right)$.
By keeping $\Delta_{\rm ads/des}$ constant and varying
$\Delta_{\rm d} < \Delta_{\rm ads/des}$, we vary the ratio
of the diffusion rate to the  adsorption/desorption rate. 
The absolute values of the barriers have no particular meaning 
unless one attempts to specify the relation between physical 
and Monte Carlo time scales. In that case they would have to be 
calculated theoretically, for instance by quantum mechanical density functional 
theory (DFT),\cite{YILD09,JUWO11}  
or by comparison between simulations and experiments.\cite{BOTT96,ABOU04} 
In the present work, we treat the barriers as merely formal quantities. 

A commonly used transition 
rate corresponding to Eq.~(\ref{eqndele}) is\cite{FRA06}
\begin{eqnarray}
R_{\rm A \rightarrow \rm B}= \nu_0 \exp\left(- \Delta/T \right )
\exp\left[- (E_{\rm B} -
E_{\rm A})/2T\right] \;,
\label{eqnrats}
\end{eqnarray} where $\nu_0^{-1}$ determines  
the Monte Carlo time scale (one Monte Carlo Step per Site, 
MCSS), which is chosen as the basic time unit in the following. 
The above rate obeys detailed balance as the ratio
$R_{A \rightarrow B}/R_{B \rightarrow A}$ cancels the prefactor,
$\nu_0 \exp\left(- \Delta/T \right )$.

The stochastic dynamics defined by this transition rate is here implemented 
by the continuous-time, rejection-free $n$-fold way Monte Carlo 
algorithm.\cite{BOR75,NOV01,GIL72,GIL76} 
This algorithm provides significant computational speed-up 
while preserving the underlying dynamics, at the expense 
of quite intricate programming. A detailed explanation of how the method is 
implemented in the model studied here is given in Ref.~\onlinecite{FRA06}.

\subsection{Measurements}
\subsubsection{Morphology}
\noindent One important measure closely related to the lateral 
diffusion is the \emph{correlation length}. 
The correlation between the occupation at two sites
$(i,j)~\mbox{and}~(i',j')$ is $c(i,j)c(i',j')$. 
(Here, the double indices, $i$ and $j$, refer to the two lattice directions.)  
With this expression, we define the \emph{correlation function} $\Gamma(l)$,
averaged over the two lattice directions, as 
\be
\Gamma (l)=\left( \frac{1}{L^2} 
     \ncs c(i,j) \frac{c(i+l,j)+c(i,j+l)}{2} \right)
- \left( \frac{1}{L^2} \ncs c(i,j) \right)^2 \;,
\label{eqngamm}
\ee
where $0~\le~l~\le~L/2$. For $l=0$, Eq.~(\ref{eqngamm}) reduces to
$\Gamma(0)=\theta(1-\theta)$, and the {\it normalized\/} 
correlation function is defined as $\gamma(l)={\Gamma (l)}/{\Gamma (0)}$.
The \emph{correlation length} $\xi$ is estimated from the inverse of the
initial slope of the normalized correlation function $\gamma(l)$.
For the discrete case of our lattice-gas model, this can be 
written as\cite{DEB57,JUW12}
\be
\xi=\frac{1}{1-\gamma(1)} \;.
\label{eqninve}
\ee
Those lattice edges that connect an occupied and an empty lattice site are 
often called ``broken bonds." Their number, $\Sigma$, 
equals the total size of the 
interface between occupied and empty sites. It can be shown that $\xi$, 
as estimated by Eq.~(\ref{eqninve}), 
is related to $\Sigma$ and $\theta$ as\cite{DEB57,BROW02,JUW12}
\begin{eqnarray}
\xi=\frac{4\theta(1-\theta)}{\Sigma/V} \;.
\label{eqndeby}
\end{eqnarray}


Other important average properties of the adsorbate configuration are the 
{\it cluster number density\/} $n_s$ and the \emph{Mean Cluster Size} $S$.  
In introducing these quantities we closely follow the exposition and notation 
of Ref.~\onlinecite{STA92}. 
For $N_s$ clusters containing $s$ occupied sites each, the cluster number 
density,  
\be
n_s={N_s}/{V} \;,
\label{eqnnclu}
\ee
is the number of such $s$-particle clusters {\it per lattice site\/}. 
Hence, the area density distribution, $n_ss$, 
is the probability that a randomly chosen site belongs to 
an $s$-cluster, and the coverage, $\theta=\sum \limits_s n_ss$,
is the probability that it belongs to {\it any\/} cluster, i.e.,
that it is occupied.
Thus, the probability that the cluster to which an arbitrary {\it occupied\/} 
site belongs contains exactly $s$ sites is
$w_s={n_ss}/{\sum \limits_s n_ss}$.
The {mean cluster size} $S$ that we measure in this process of randomly 
hitting some occupied site is therefore
\be
{S} = \sum \limits_s w_ss  
    = \frac{\sum \limits_s n_ss^2}{\sum \limits_s n_ss} \;.
\label{eqnmean}
\ee
The magnitude of change of size distributions at a given coverage is measured
by the changes in the {mean cluster size} $S$
and \emph{the cluster density},
\begin{eqnarray}
\rho_\delta = \sum_s n_s \;,
\label{eqndrop}
\end{eqnarray}
which is the total number of clusters per unit area.

The details of the size distribution at the beginning of and 
during the desorption are obtained from number density ($n_s$) 
histogram plots,\cite{FRA05} each averaged over 100 independent runs. 
Therefore, $n_s$ becomes an estimate for the probability of finding the
center of mass of a cluster of size $s$ at a randomly chosen 
site.\cite{FRA05,STA92}
The number density for larger clusters is very low compared to that of 
the small clusters. 
For that reason  we use exponentially growing bins  for the histograms. 
The bins are set up
such that each bin is twice as large as the previous one.
This results in an uniform distribution of data points on a 
logarithmic scale.\cite{Norms}

\subsubsection{Dynamics}
\label{secYYY}
\begin{figure}[h!]
    {\includegraphics[scale=0.6]{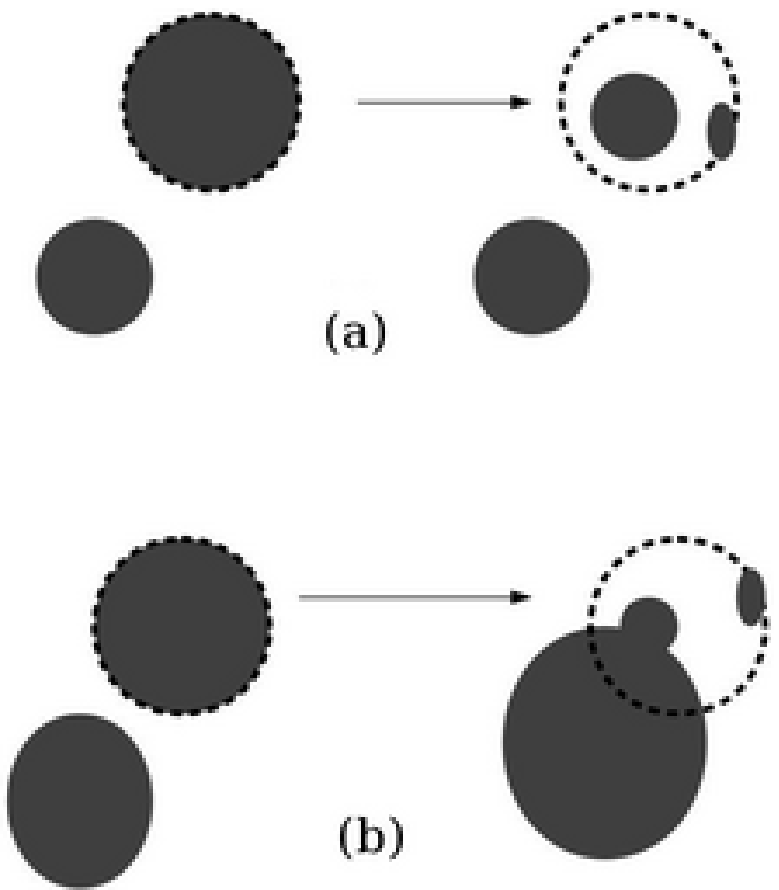}}
\caption{Cluster tagging. (a) A cluster shrinks and splits into two fragments.
Here the larger of the two is taken as representative of the cluster.
 (b) A cluster shrinks and splits into two multi-atom fragments. 
One of these fragments coalesces with a
larger cluster. The new coalesced cluster is then taken as representative 
of the split cluster.}
\label{pictag1}
\end{figure}

The dynamics of the entire system is studied by measuring $\theta(t)$, while
the individual cluster dynamics are observed by tagging a specific cluster 
at the beginning of the desorption process, and following it while
measuring its size as a function of time, $s_i(t)$, during the whole process.
Figure~\ref{pictag1} illustrates our cluster-tagging procedure. 
We pick a specific cluster at the beginning of the desorption simulation and
record the coordinates of all the points in the cluster. During
the simulation process we monitor those coordinates. We record
the cluster labels at each of those coordinates at any given time and
obtain the cluster size as a function of time. Figure~\ref{pictag1}(a) gives
 an example in which the cluster shrinks and splits into two fragments.
In this case we take the largest cluster  within the circle as 
the representative of the cluster. There is also a possibility
that the cluster just splits into two fragments of comparable
sizes. In this case, a sudden drop in the cluster size is seen when 
it is plotted as a function of time.
Figure~\ref{pictag1}(b) shows
a different case, in which the original cluster coalesces with another
cluster and becomes a new, larger cluster. Since part of the new cluster
is still within the circle, we take the new, coalesced cluster as the representative.
Here  a sudden jump in the cluster size is seen when it is plotted 
as a function of time. 
There is also a possibility that the original cluster simply disappears
from the circle and is consumed by a larger cluster. In this case, we
label the event as cluster disappearance. 
Results for the time evolution of some representative clusters are discussed in 
Sec.~\ref{secCS}. 

\section{Simulation Preparation}
\label{preparation}

\noindent We start from an empty lattice and equilibrate the system at negative 
electrochemical potential, $\mu-\mu_0=-0.4$, 
to achieve a very low coverage before
switching on a positive potential until a coverage  
cutoff $\theta_{\rm cutoff}$ is reached. The configuration at 
$\theta_{\rm cutoff}$ then
becomes the initial configuration for the desorption processes with
$\theta_{\rm init}=\theta_{\rm cutoff}$. We prepare a set of four classes of 
initial 
configurations by applying four different electrochemical potentials 
$\mu-\mu_0=0.4,~1.6,~2.56,~{\rm and}~9.76$ during the adsorption process. 
The average cluster size at $\theta_{\rm init}$ decreases
with increasing $\mu-\mu_{\rm 0}$.
The coverage cutoff is chosen to be $\theta_{\rm init}=0.35$ in all cases.
This is larger than the maximum coverages used for the 
experiments reported in Refs.\ \onlinecite{HEY01,BAR95,BAR96}. 
However, at the temperature used in this study, the equilibrium configuration 
at a fixed coverage between 0.35 and about 0.1 is expected to be a 
single monolayer cluster surrounded by a low-density gas of monomers and 
much smaller clusters. No giant or percolating adsorbate 
cluster is expected, even at 
equilibrium.\cite{JLEE95,BIND03,BISK03,NUSS08}. 
We consider this sufficient 
to ensure the qualitative relevance of our results to such experiments. 

For each run of the simulation, we performed adsorption until 
$\theta_{\rm cutoff}$ was reached,  
immediately followed by desorption until a low-coverage 
equilibrium was re-established. During the adsorption stage, the 
adsorption/desorption barrier was fixed at $\Delta_{\rm ads/des}=15$. 

For the desorption stage, we fixed the potential at $\mu-\mu_0=-0.4$, 
and we also kept the adsorption/desorption barrier unchanged at 
$\Delta_{\rm ads/des}=15$. To study the effects of diffusion, we performed two 
different desorption runs for each initial configuration;  one effectively 
without diffusion 
(for convenience implemented simply by setting $\Delta_{\rm d}=150$), 
and the other with relatively fast diffusion ($\Delta_{\rm d}=8$). 

\begin{figure}[!ht]
\begin{center}
$$
\begin{array}{cc}
    {\rm (A) } & {\rm (B)} \\
    {\includegraphics[scale=0.8]{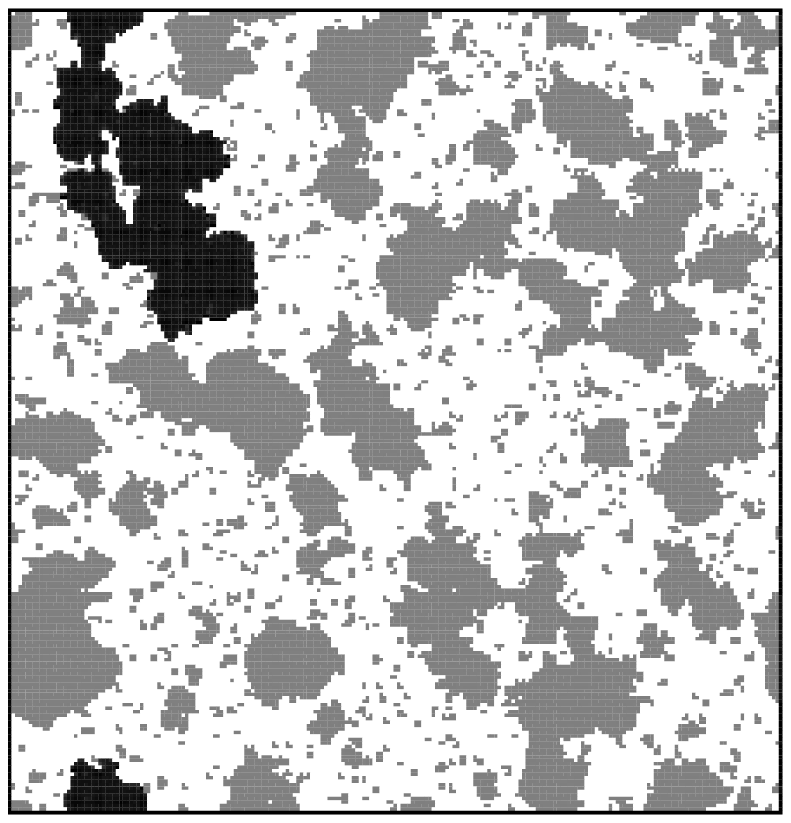}} &
    {\includegraphics[scale=0.8]{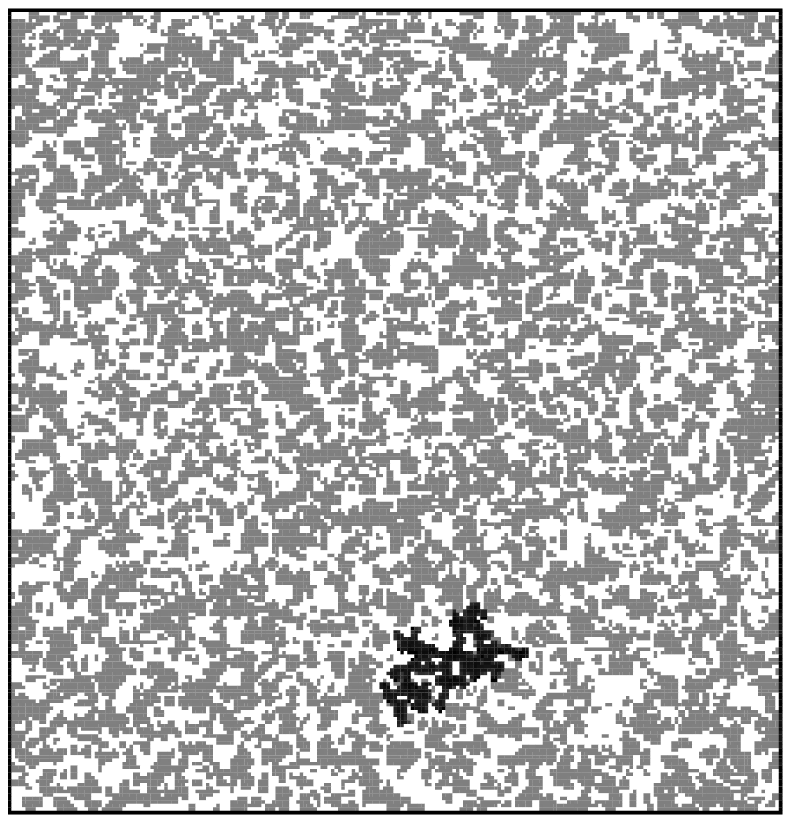}} \\
    {\rm (C) } & {\rm (D)} \\
    {\includegraphics[scale=0.8]{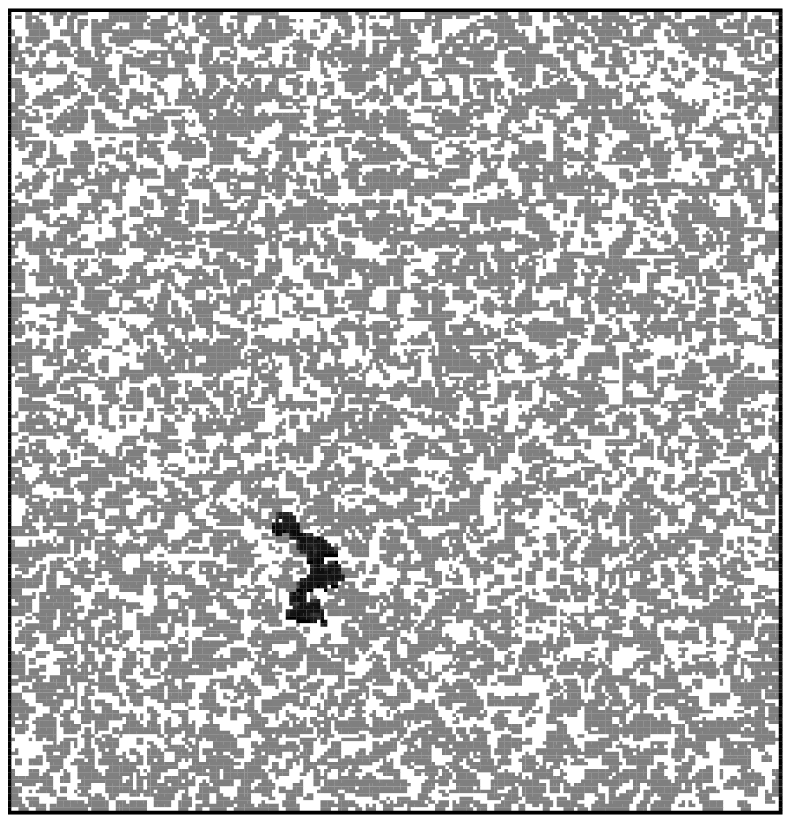}} &
    {\includegraphics[scale=0.8]{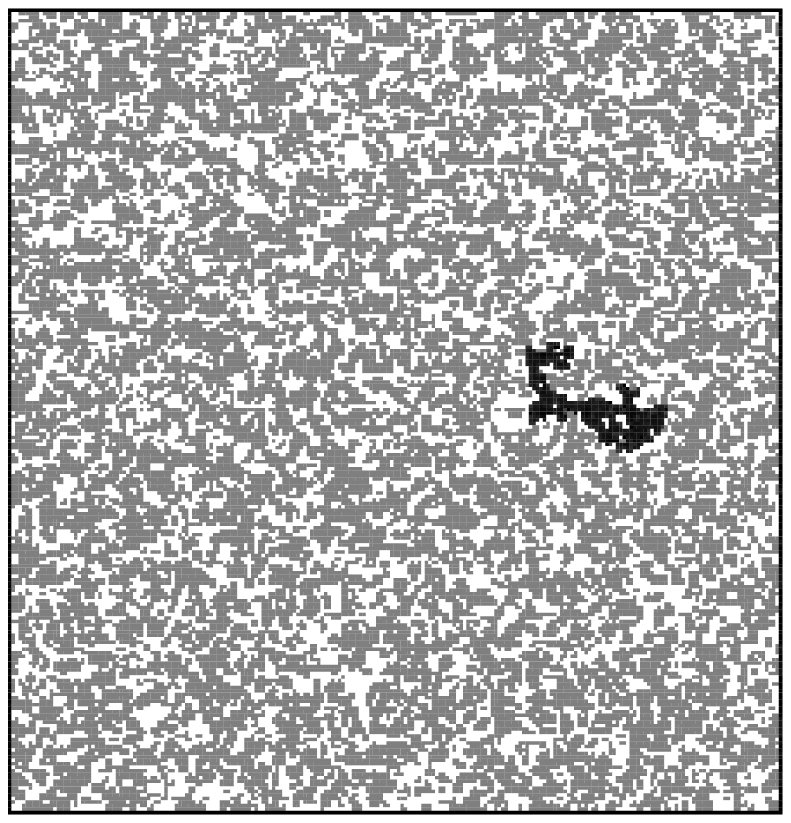}} \\
\end{array}
$$
\end{center}
\caption{Typical snapshots of the initial configurations of the
four desorption simulations at $\theta_{\rm init}=0.35$, 
with $\xi_{\rm init}$ = (A) 4.96,
 (B) 2.35, (C) 1.96, and (D) 1.55, respectively. 
 The largest cluster in each picture is colored dark.}
\label{DROPLET1}
\end{figure}
\begin{figure}[!ht]
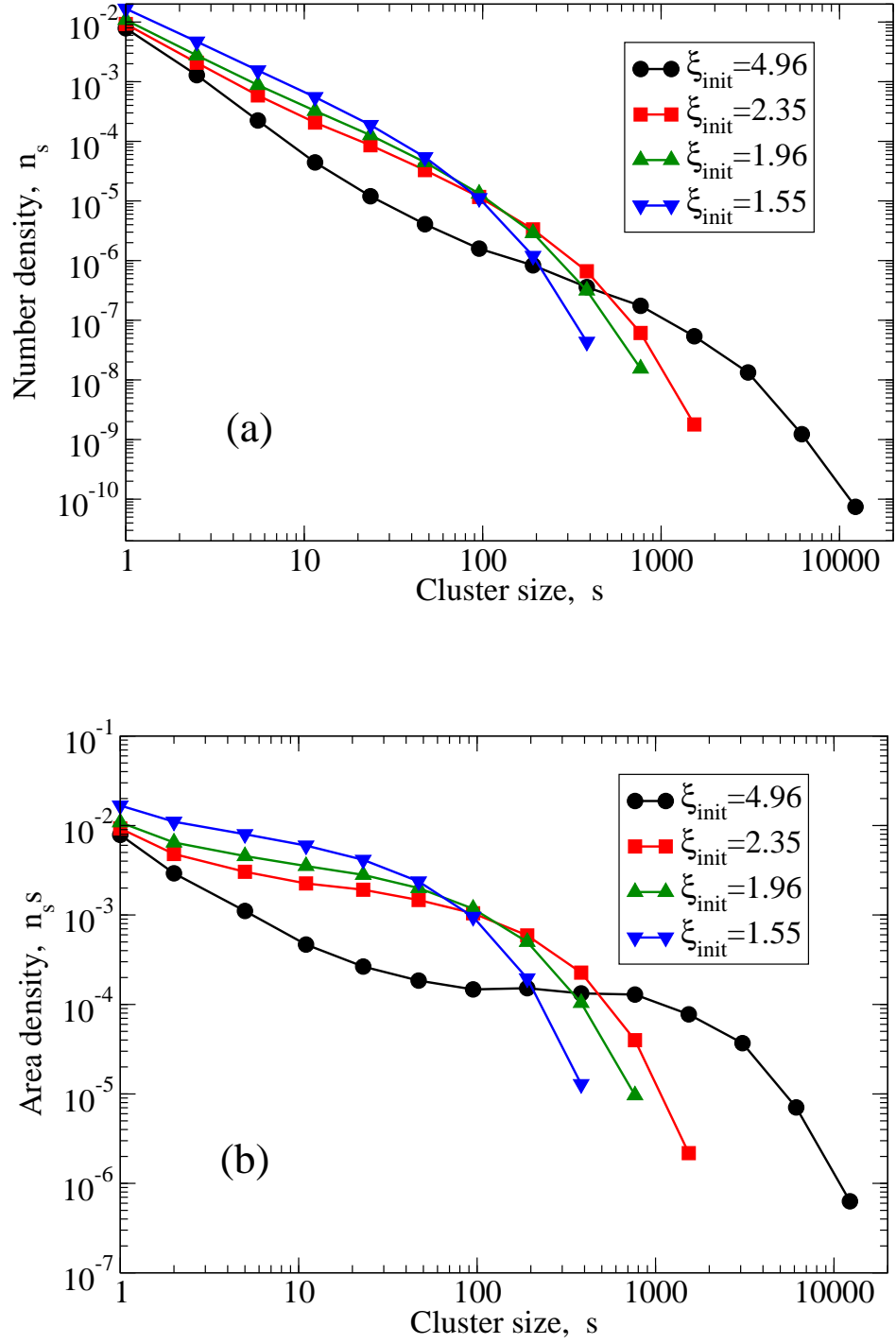

\begin{center}
    {\includegraphics[scale=0.5]{init_per.eps}}
\vskip 1.5truecm
    {\includegraphics[scale=0.5]{NSS1.eps}}
\end{center}
\caption{(a) Initial cluster size distributions, $n_s$ vs $s$, 
for the four simulations $\xi_{\rm init}$= (A) 4.96 (circles), 
(B) 2.35 (squares), (C) 1.96 (triangles up), and (D) 1.55 (triangles down). 
All are at coverage $\theta_{\rm init}=0.35$.
(b) Corresponding data for the area density distributions, 
$n_s s$ vs $s$.\protect\cite{Norms}
}
\label{Initconf}
\end{figure}

Figure~\ref{DROPLET1} shows typical snapshots for each of the four initial 
configurations at $\theta_{\rm init}=\theta_{\rm cutoff}$. 
For convenience, we label each class of initial configurations 
and each of the corresponding sets of 
desorption simulations as (A), (B), (C),  and (D). 
Visual inspection of the snapshots immediately shows that configuration~(A) 
is dominated by large clusters, while (B), (C), and (D) 
are dominated by progressively 
 smaller clusters, respectively. This is confirmed by 
measurement of the mean cluster size $S$ for each of the configurations. 
Table~\ref{TAB01} summarizes the parameters of the four initial configurations. 
The cluster number densities, $n_s$, of each of the four classes of 
initial configurations are 
shown in Fig.~\ref{Initconf}(a), and the corresponding  
area densities, $n_s s$, in Fig.~\ref{Initconf}(b). From Table~\ref{TAB01}, 
it is evident that reducing the mean cluster size results in reducing 
the correlation length as well.  
In our case, the four initial size distributions 
(A), (B), (C), and (D) show some
similarity. The number density  vs cluster size is
always monotonically decreasing, showing a broad hierarchy of sizes. 
In configuration (A), the numbers of medium and small clusters are smaller, compared to the other
configurations. 
The same can be said when we compare (B) to (C) and (C) to (D). 
However, the number density of larger clusters is smaller than the number density of smaller clusters in all of these four cases.

\section{Simulation Results}
\label{results}

\subsection{Coverage, Correlation Length, and Cluster Density}
\label{secCC}

Our measurements are focused on the changes in morphology
and dynamics due to the lateral diffusion as the initial configurations 
are varied.
Table~\ref{TAB03} summarizes the simulation results in terms of
those changes for several coverages between $\theta_{\rm init} = 0.35$ and 
$\theta = 0.05$. The quantities measured are the increase of the 
correlation length, $\Delta \xi= \xi'-\xi$, 
the fractional increase of the mean cluster size, $\delta S=(S'-S)/S$, 
and the decrease in the desorption time required to reach a given coverage, 
$\Delta t(\theta)=t(\theta)-t'(\theta)$ 
between the simulations without and with diffusion. 
The unprimed quantities in these expressions indicate the simulations
without diffusion,
while primes refer to the simulations with diffusion.

For $\theta=$ 0.32, 0.28, 0.25, 0.18, and 0.12, $\Delta \xi$ is always 
positive and increases
as we decrease the initial correlation length from $\xi_{\rm init}$ = 4.96
 to 2.35, 1.96, and 1.55, respectively. 
The coverage $\theta=0.05$ is close to the equilibrium value, in which 
the measurement results are small fluctuations around a constant value.
At the earliest times (i.e., coverages closely below $\theta_{\rm init}$), 
the presence of diffusion {\it retards\/} the desorption process. This is 
indicated in Table~\ref{TAB03} 
by the negative values of $\Delta t$ in simulations (B)-(C). 
Later in the simulations, the retardation is replaced by {\it accelerated\/} 
desorption, indicated by positive $\Delta t$. 

In Fig.~\ref{INSET1} we show the time evolution of the coverage $\theta$,  
with   insets  showing corresponding data for the correlation length $\xi$.  
The main parts of the figure confirm that diffusion induces 
a {\it crossover\/} from retarded 
desorption at early times, to accelerated desorption at late times. In part 
(A), corresponding to the large $\xi_{\rm init} = 4.96$, 
the crossover to acceleration occurs almost immediately after the start of 
the desorption. [Thus deceleration 
is not seen for simulation (A) at the coverages included in Table~\ref{TAB03}.] 
Conversely, in part (D), corresponding to the small $\xi_{\rm init} = 1.55$, 
the desorption remains retarded except at the very latest times. For the 
intermediate values of $\xi_{\rm init}$ shown in parts (B) and (C), the 
crossover can be clearly seen at an intermediate time. 
The insets, showing the effects of diffusion on the correlation length
$\xi(t)$, display an analogous crossover from {\it coarsening\/} 
at early times to a reduction of $\xi$ at late times. 
We explain these competing effects of diffusion during the desorption 
process as follows. 

{\it Retardation.\/}
Due to their lack of lateral bonding, monomers are the most easily desorbed 
particles. However, lateral diffusion provides a mechanism for them to 
move into contact and bond with larger clusters or other monomers, thus 
reducing their subsequent rate of desorption. This process corresponds to 
a coarsening of the adsorbate configuration since it reduces the number of 
broken bonds [$\Sigma$ in Eq.~(\ref{eqndeby})] by between two and six, 
without changing $\theta$.

{\it Acceleration.\/}
Conversely, diffusion 
also provides a mechanism for a particle at the surface of a 
cluster to diffuse away and become a monomer. Following its 
detachment from the cluster, the new monomer has a higher desorption rate. 
While the initial detachment by diffusion clearly increases $\Sigma$, a 
following desorption of the resulting monomer will reduce $\Sigma$ by four and 
$\theta$ by $1/V$. The total change in $\xi$ may therefore be positive or 
negative, depending on the values of both these variables.  

The balance between deceleration/coarsening and acceleration is determined 
by the  fraction of the total coverage that consists of monomers and small 
clusters. This fraction is small for large $\xi_{\rm init}$. As a result, 
acceleration dominates as monomers ``evaporate" from the large clusters. 
For small $\xi_{\rm init}$ the fraction of small clusters is large, leading to 
coarsening and deceleration of the desorption as excess monomers diffuse to 
join larger, more stable clusters. We discuss these effects in more detail 
below, in Sec.~\ref{secCS}. 



A complementary view of the effects of diffusion is provided by 
plotting the correlation length
as a function of coverage, $\xi(\theta)$, as shown in Fig.~\ref{Cov1}.
By hiding the effects of diffusion on the speed of desorption, this 
enables us to observe its influence on the adsorbate morphology 
as expressed by the correlation length at a specific coverage. Coarsening is 
seen at all coverages. The extent of the coarsening is very small 
for the largest $\xi_{\rm init}$, but increases gradually with decreasing 
$\xi_{\rm init}$. This is a reasonable result, since this view essentially 
represents a mapping of the adsorbate configuration onto a snapshot of
one that could be produced due to Ostwald ripening by 
lateral monomer diffusion at fixed coverage (Kawasaki dynamics). 
Smaller $\xi_{\rm init}$ would then correspond to earlier times.

\begin{figure}[!ht]
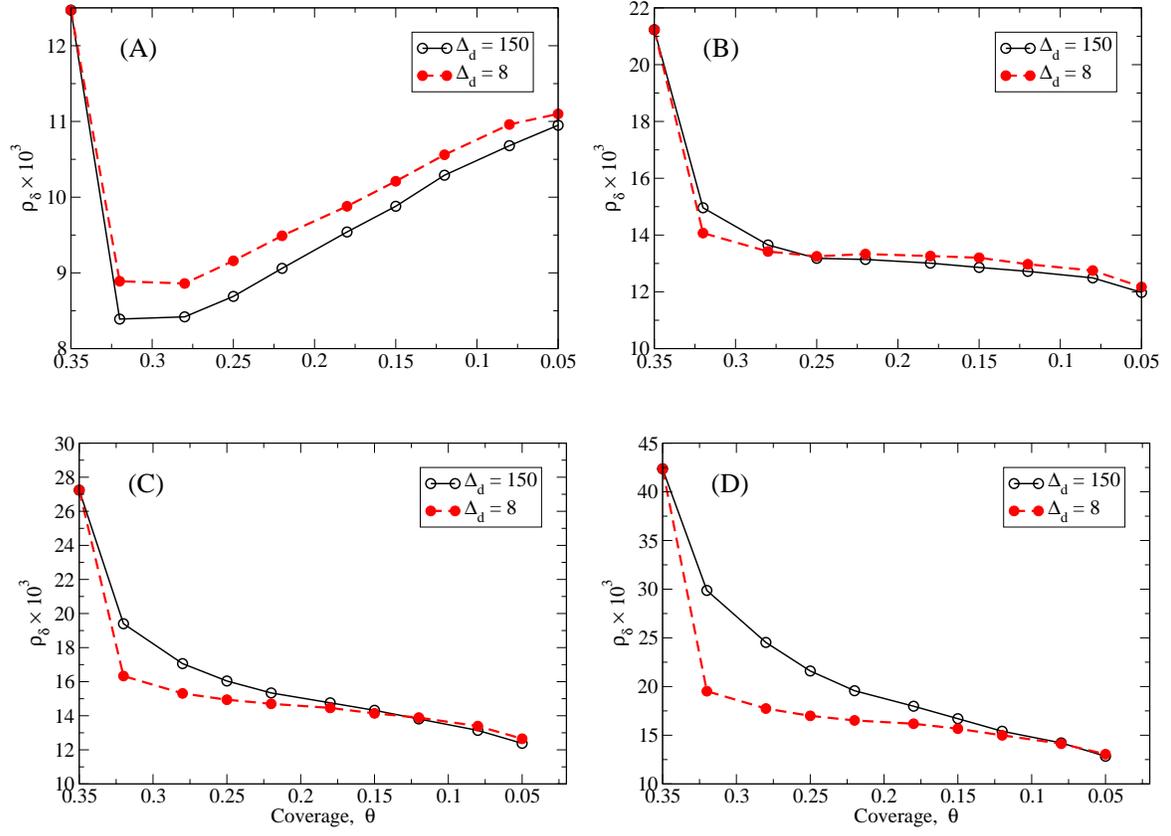

\begin{center}
$$
\begin{array}{cc}
    {\includegraphics[scale=0.3]{DD1.eps}} &
    {\includegraphics[scale=0.3]{DD2.eps}} \\
    { } & { } \\
    {\includegraphics[scale=0.3]{DD3.eps}} &
    {\includegraphics[scale=0.3]{DD4.eps}} \\
\end{array}
$$
\end{center}
\caption{Cluster density $\rho_\delta(\theta)$ for the 
simulations with $\xi_{\rm init}$= (A) 4.96, 
(B) 2.35, (C) 1.96, and (D) 1.55. Open circles with solid lines 
represent simulations without diffusion, 
and full circles with dashed lines represent simulations
with diffusion. Note
 that $\theta$ {\it decreases} toward the right in this figure.}
\label{RHO1}
\end{figure}

The effect of diffusion during the early stages
of desorption is connected to the average distance between  
clusters. At a given coverage, this average distance is 
inversely proportional to the cluster density $\rho_{\delta}$. 
Figure~\ref{RHO1} 
shows the results of cluster density measurements 
$\rho_{\delta}(\theta)$. Initially, the cluster density
quickly drops by between 
30\% and 50\% as $\theta$ decreases from its initial value. 
In simulation (A), it increases 
again after the initial drop, while in simulations (B), (C), and (D), it
continues to decrease at a lower rate.

The effect of diffusion on the cluster density depends on the initial 
correlation length. 
In simulation (A),  diffusion increases $\rho_{\delta}$  
for all values of $\theta$ included in the figure. 
For the smaller values of $\xi_{\rm init}$ (B-D) there is a crossover from 
a reduction of $\rho_{\delta}$ for larger $\theta$ (early times) to an 
increase for smaller $\theta$ (later times). Comparing the coverages at which 
this crossover occurs with the plots of $\theta$ vs $t$ in 
Fig.~\ref{INSET1}, we see that they correspond approximately to the crossover 
times seen in that figure. Thus, a reduction of $\rho_{\delta}$ due to 
diffusion is correlated with retarded desorption, while an increase in 
$\rho_{\delta}$ is correlated with accelerated desorption.

\subsection{Cluster Size Distributions and Behavior of Individual Clusters}
\label{secCS}

The morphological changes by diffusion
 are further illustrated by the area density histograms shown
in Fig.~\ref{32HIS1}. These figures show an example of 
the overall changes in $n_s s$ by diffusion
at a particular coverage, $\theta=0.25$. 
Figure~\ref{32HIS1}(A) shows a small change by diffusion,  
consistent with the very small increase of $\xi(\theta)$ by diffusion 
shown in Fig.~\ref{Cov1}(A). Figures \ref{32HIS1}(B-D) show 
more significant changes, 
consistent with larger  changes in $\xi(\theta)$ by diffusion
as we reduce  $\xi_{\rm init}$. The changes amount to  
a depletion of intermediate-size clusters, 
which is partially compensated by transfer of coverage to larger clusters and 
monomers. This effect is observed for all coverages below $\theta_{\rm init}$. 

Next we turn our attention to the dynamics of individual clusters.
In Fig.~\ref{Shrink1} we show examples of the time evolution of the 
size $s$ of four individual, 
relatively large clusters picked from different simulations. 
The results were obtained with the cluster-tagging method discussed 
in Sec.~\ref{secYYY}. We see a combination of shrinkage, occasional growth, 
splitting, and coalescence. Diffusion is seen to increase the 
frequencies of both coalescence and splitting events.

Figure~\ref{TLargest1} shows the average time evolution of the largest 
clusters at the beginning of the desorption stage (dark-colored clusters 
in Fig.~\ref{DROPLET1}), each picked out
from realizations of the four classes of initial conditions (A--D),  
and each averaged over 100 independent runs.
We immediately notice a qualitative similarity of these four average 
cluster behaviors to the general behavior of $\theta(t)$ in 
Fig.~\ref{INSET1}, including the crossover from retarded to accelerated 
desorption. 

\begin{figure}[!ht]
\begin{center}
{\includegraphics[scale=0.55]{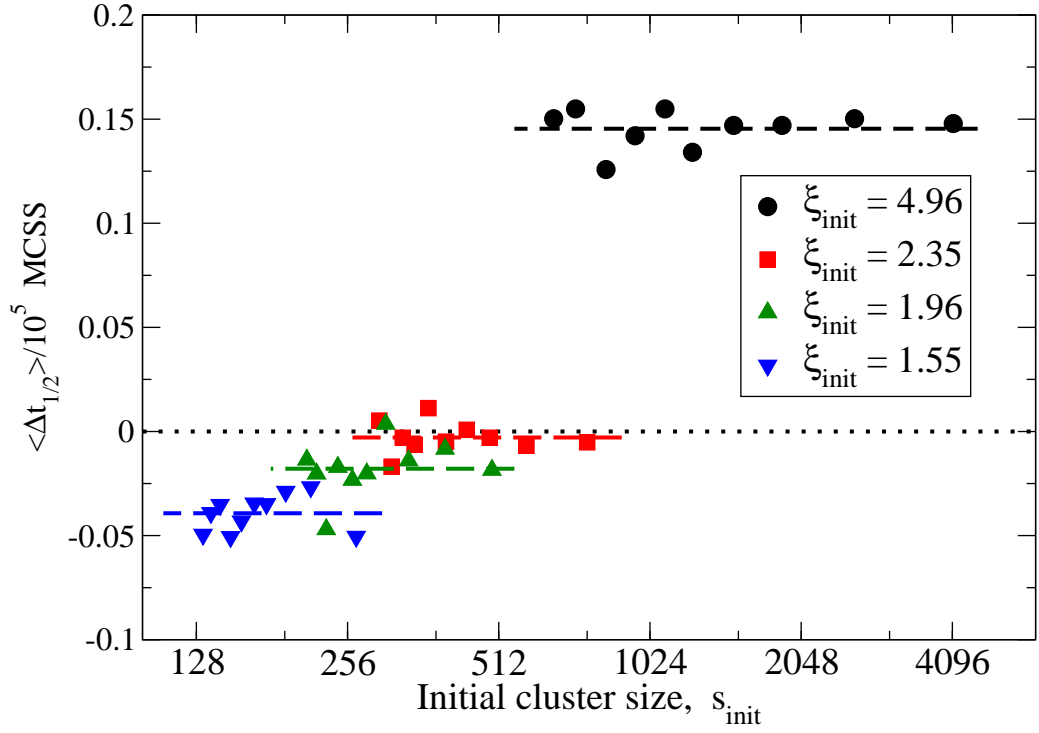}}
\end{center}
\caption{Halftime decrease with diffusion for the
10 largest clusters taken from a single realization of the initial conditions,
$\xi_{\rm init} = $ (A) 4.96 (circles), 
(B) 2.35 (squares), (C) 1.96 (triangles up), and (D) 1.55 (triangles down). 
Each result is averaged over 100 independent desorption runs. 
Positive (negative) results correspond to acceleration (deceleration). 
The horizontal dashed lines represent the averages corresponding to each of the 
four data sets. 
}
\label{sv1}
\end{figure}

To further investigate the dynamics of the individual clusters, we
pick out the 10 largest clusters 
from one realization of each of the four initial conditions (A-D)
and measure the time to reach half their original volume 
(halftime $t_{\rm 1/2}$). 
Each result is averaged over 100 independent desorption runs. We then
measure the acceleration,
\begin{equation}
\Delta t_{\rm 1/2} = t_{\rm 1/2} -t'_{\rm 1/2}.
\label{eqdelt}
\end{equation}
Figure~\ref{sv1} shows the decrease of the halftimes (positive for 
acceleration). 
We observe a crossover from acceleration to deceleration with decreasing 
$\xi_{\rm init}$. Comparing with the data for the {\it total\/} coverage 
$\theta$ vs $t$ in 
Fig.~\ref{INSET1}, we see that the behavior of the largest 
clusters in a single initial configuration 
is predictive of the overall acceleration/deceleration behavior of many 
initial configurations with the same $\xi_{\rm init}$
around the time when the average 
{total} coverage reaches half its initial value.

\section{Conclusions}
\label{analysis}

Inspired by experimental results,\cite{HEY01,BAR95,BAR96} 
we studied the effects of lateral monomer 
diffusion on a lattice-gas model for the desorption of a submonolayer of 
adsorbate atoms from a single-crystal surface. 
Using large-scale kinetic Monte Carlo simulations, we found that diffusion 
produces two competing effects, and that the balance between the two 
depends on the shape of the cluster-size distribution at the onset 
of desorption.  

{\it Retardation\/} of the desorption process is associated with the diffusion 
of monomers, which are weakly bound and thus easily desorbed, to achieve more 
strongly bound positions as members of multiatom clusters. This leads to a 
coarsening of the adsorbate configuration, expressed by an increase in the 
adsorbate correlation length $\xi$ and a depletion of the population of 
intermediate-size clusters. This effect is dominant when the adsorbate 
configuration is dominated by monomers and small clusters. 

{\it Acceleration\/} of the desorption process is associated with the 
diffusion-induced ``evaporation" of monomers from the perimeters of larger 
clusters to the surrounding two-dimensional, low-density 
``adsorbate gas," where the absence of lateral 
bonding enhances their desorption rate. This effect is dominant when 
the adsorbate configuration is dominated by larger clusters. In 
contrast to the retardation, this mechanism can lead to an increase or a 
decrease in the correlation length, depending on the coverage and the 
number of broken bonds. 

\begin{figure}[!ht]
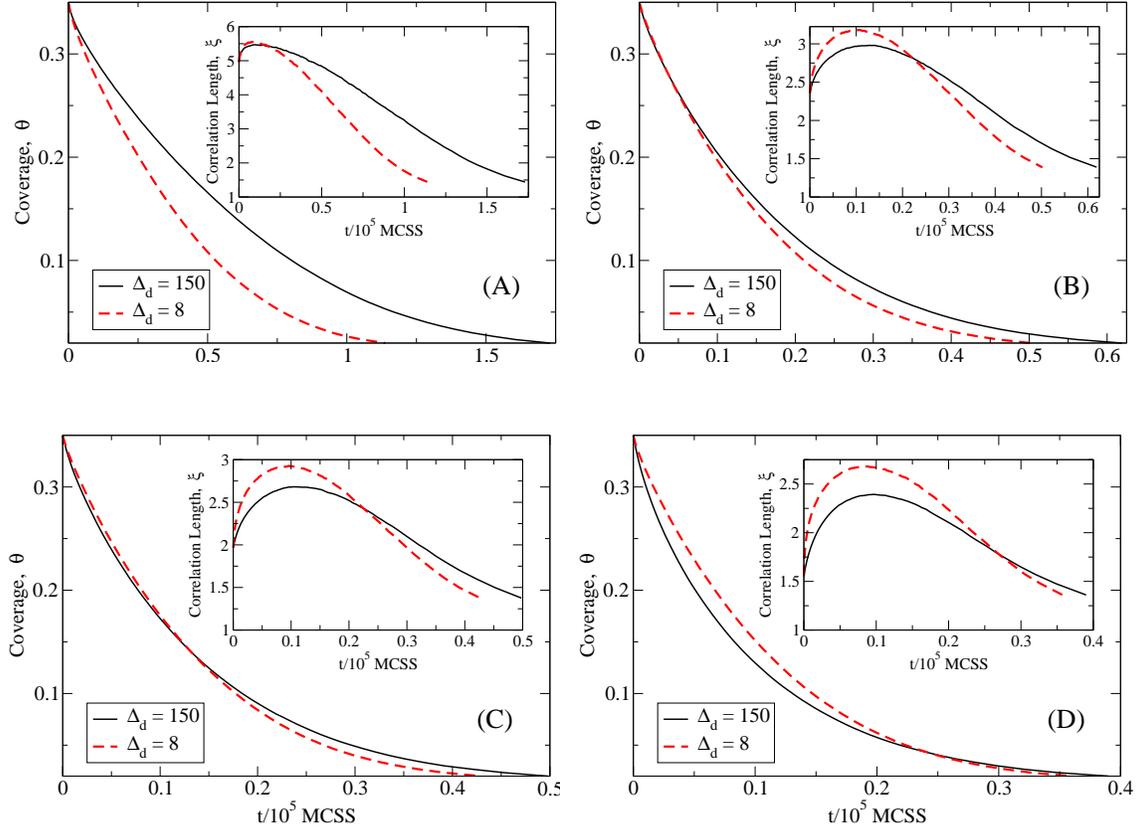

\begin{center}
$$
\begin{array}{cc}
    {\includegraphics[scale=0.3]{INSET1.eps}} &
    {\includegraphics[scale=0.3]{INSET2.eps}} \\
    { } & { } \\
    {\includegraphics[scale=0.3]{INSET3.eps}} &
    {\includegraphics[scale=0.3]{INSET4.eps}} \\
\end{array}
$$
\end{center}
\caption{The time evolution of $\theta(t)$ and $\xi(t)$ (insets) without
and with diffusion, for simulations  with
$\xi_{\rm init}$ = (A) 4.96, (B) 2.35, (C) 1.96,  and (D) 1.55, 
each averaged over 100 runs.
 Solid
lines represent simulations without diffusion, and dashed lines represent
simulations with diffusion.}
\label{INSET1}
\end{figure}

\begin{figure}[!ht]
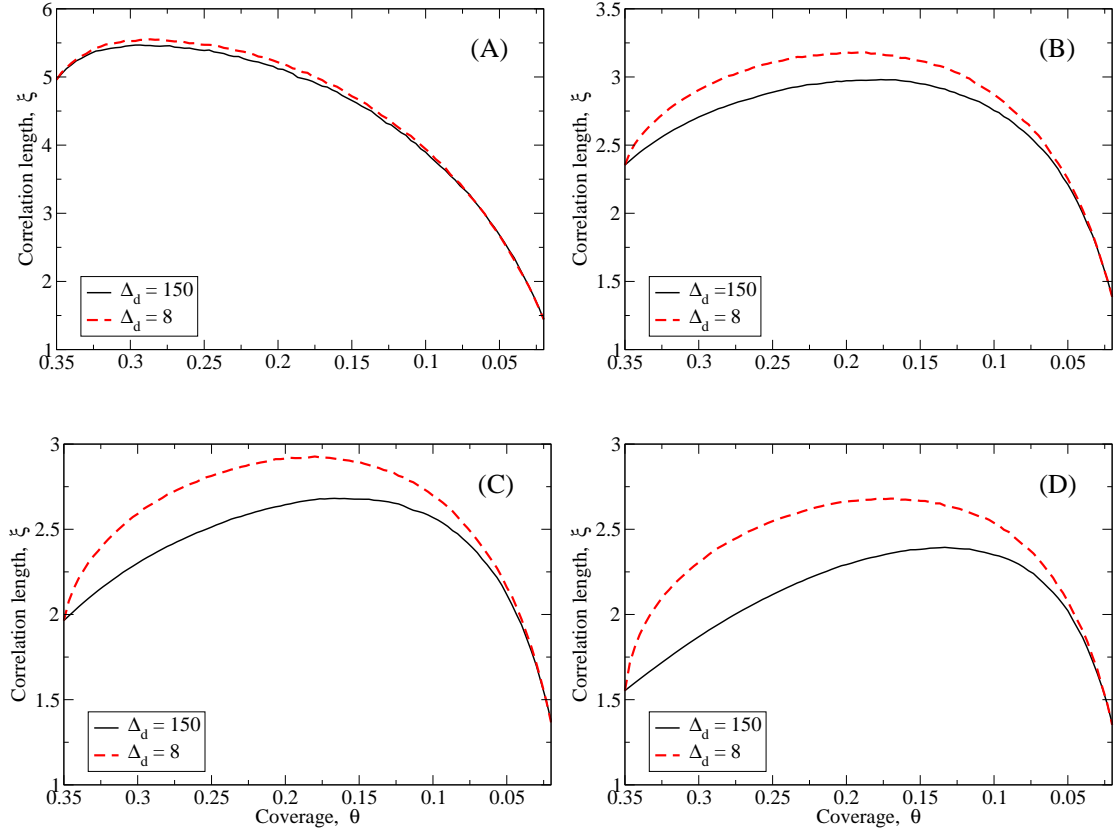

\begin{center}
$$
\begin{array}{cc}
    {\includegraphics[scale=0.3]{CorCov1.eps}} &
    {\includegraphics[scale=0.3]{CorCov2.eps}} \\
    { } & { } \\
    {\includegraphics[scale=0.3]{CorCov3.eps}} &
    {\includegraphics[scale=0.3]{CorCov4.eps}} \\
\end{array}
$$
\end{center}
\caption{The correlation length $\xi(\theta)$, 
without and with diffusion for the four
simulations with  $\xi_{\rm init}$= (A) 4.96, 
(B) 2.35, (C) 1.96, and (D) 1.55. Solid lines for simulations without
diffusion, and dashed lines for simulations with diffusion. Note
 that $\theta$ {\it decreases} toward the right in this figure.}
\label{Cov1}
\end{figure}

\begin{figure}[!ht]
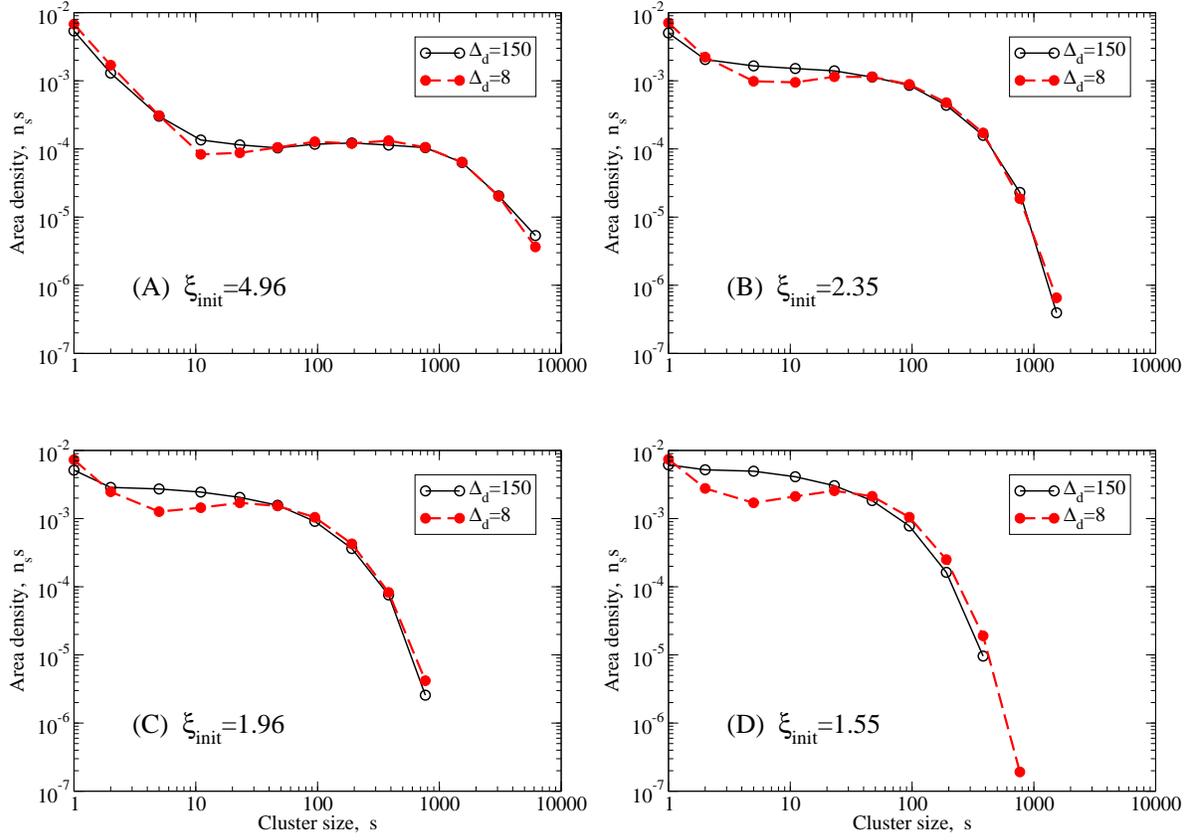

\begin{center}
$$
\begin{array}{cc}
    {\includegraphics[scale=0.3]{Nss1Y.eps}} &
    {\includegraphics[scale=0.3]{Nss2Y.eps}} \\
\\
    {\includegraphics[scale=0.3]{Nss3Y.eps}} &
    {\includegraphics[scale=0.3]{Nss4Y.eps}} \\
\end{array}
$$
\end{center}
\caption{Area density distributions $n_s s$ vs $s$ at 
$\theta=0.25$ for simulations with  $\xi_{\rm init}$= (A) 4.96, 
(B) 2.35, (C) 1.96, and (D) 1.55. Open circles with solid lines 
represent simulations without diffusion, 
and full circles with dashed lines represent simulations with 
diffusion.\protect\cite{Norms}
}
\label{32HIS1}
\end{figure}

The crossover from retardation at early times to acceleration at later times 
occurs at a time that 
depends on the initial adsorbate configuration as shown in 
Fig.~\ref{INSET1}. This is a central result of our study. A complementary 
view is provided in 
Fig.~\ref{Cov1}, which shows the correlation length vs the total adsorbate 
coverage.  Coarsening is observed at all coverages, but for a 
given coverage it is more pronounced when the initial configuration is 
dominated by monomers and small clusters. By ``filtering out" the effects of 
diffusion on the overall desorption rate, this view emphasizes the 
Ostwald ripening driven by the coverage-conserving diffusion 
process. The corresponding transfer of coverage from intermediate-sized 
clusters to monomers and large clusters is illustrated in Fig.~\ref{32HIS1}.  
The effect is strongest when the initial configuration is 
dominated by the smallest clusters. 

\begin{figure}[ht]
\begin{center}
$$
\begin{array}{cccc}
    {\rm (a)} & {\rm (b)} \\
    {\includegraphics[scale=0.6]{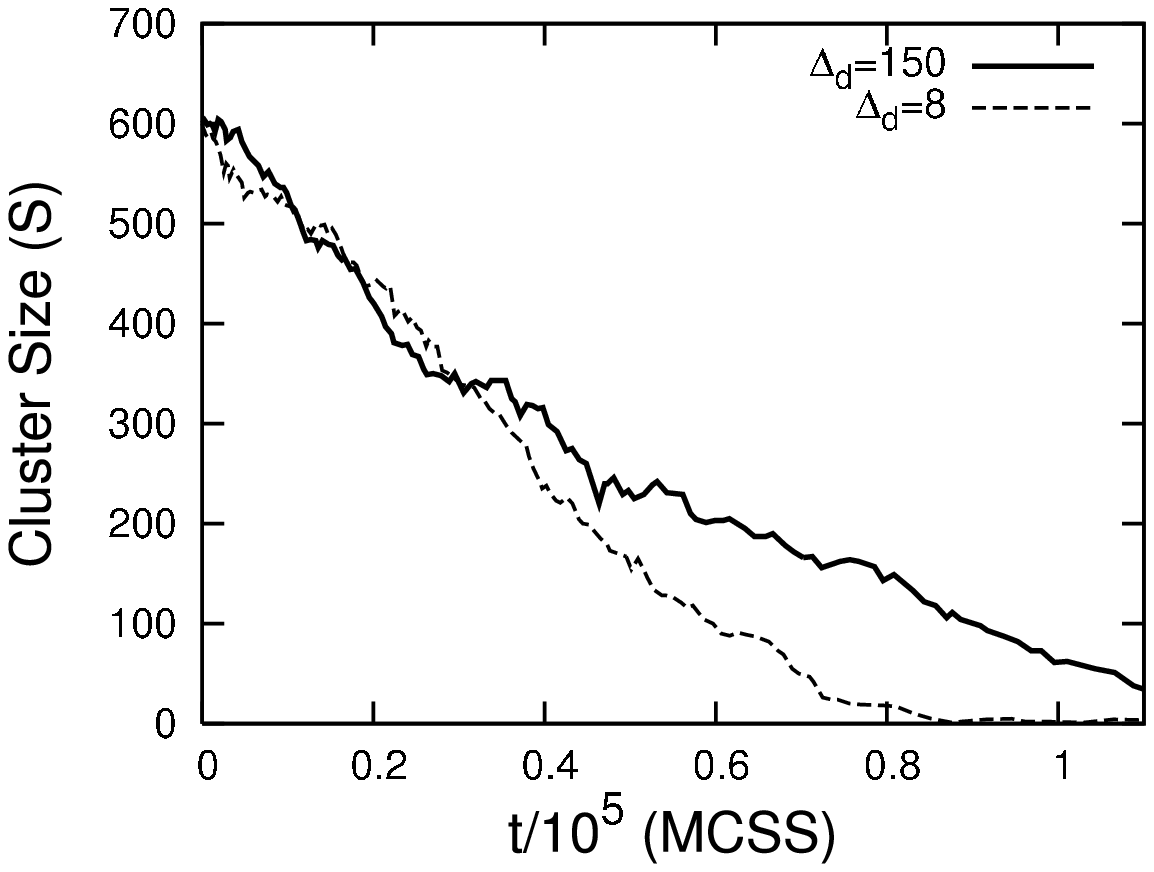}} &
    {\includegraphics[scale=0.6]{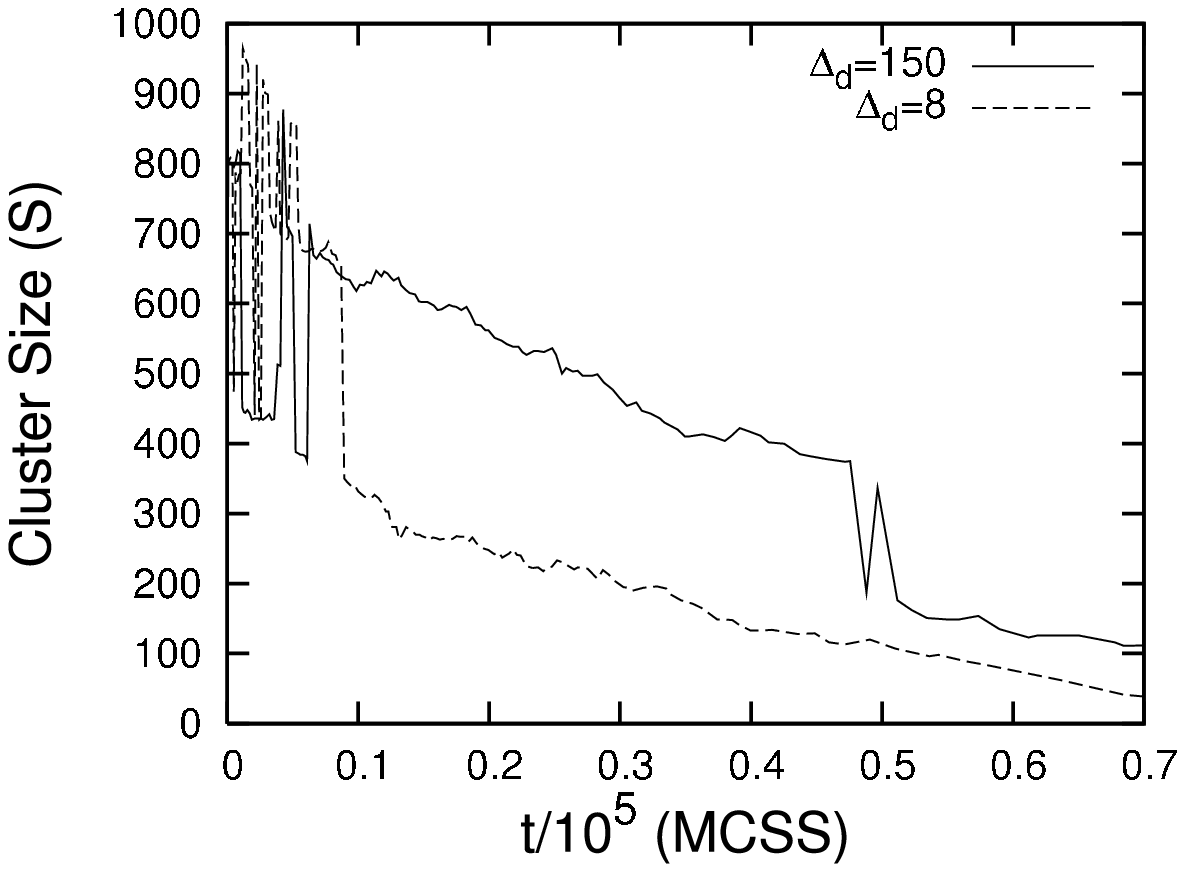}} \\
    {\rm (c)} & {\rm (d)} \\
    {\includegraphics[scale=0.6]{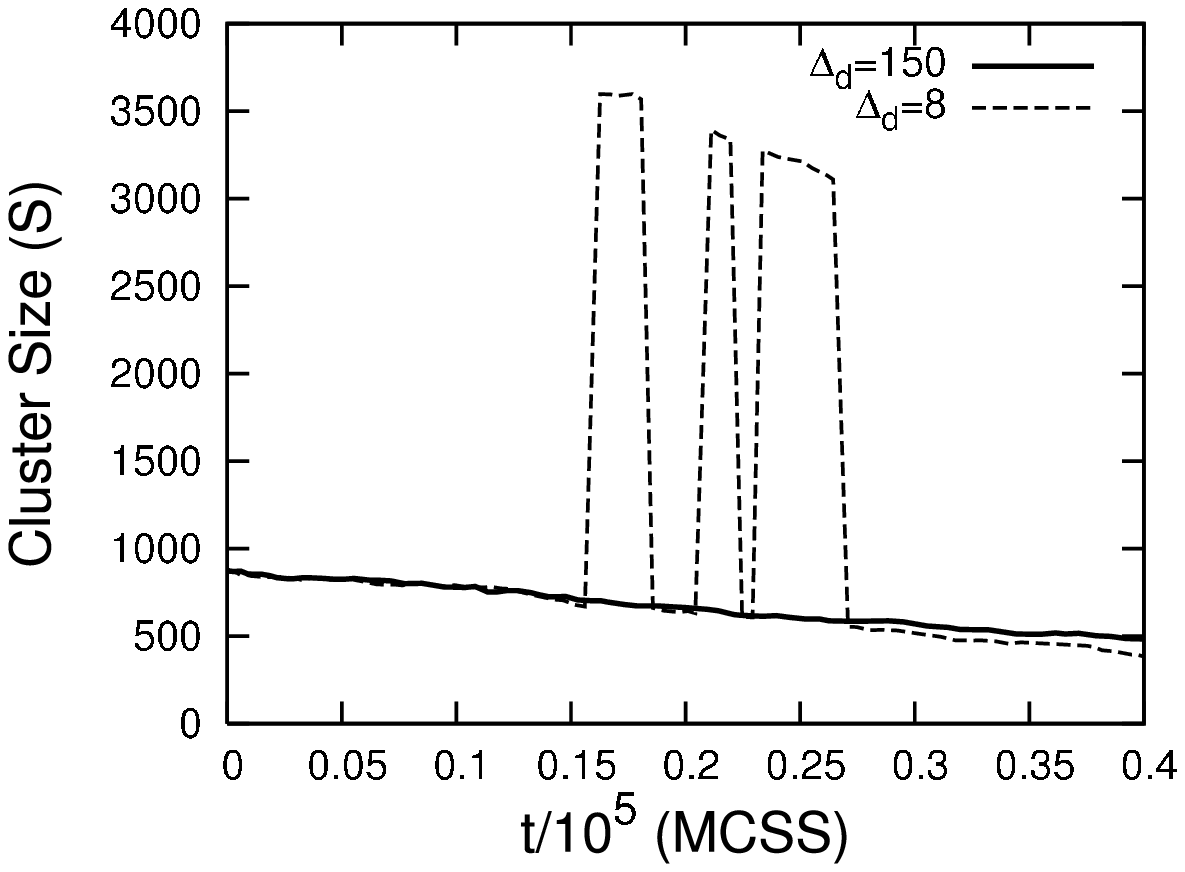}} &
    {\includegraphics[scale=0.6]{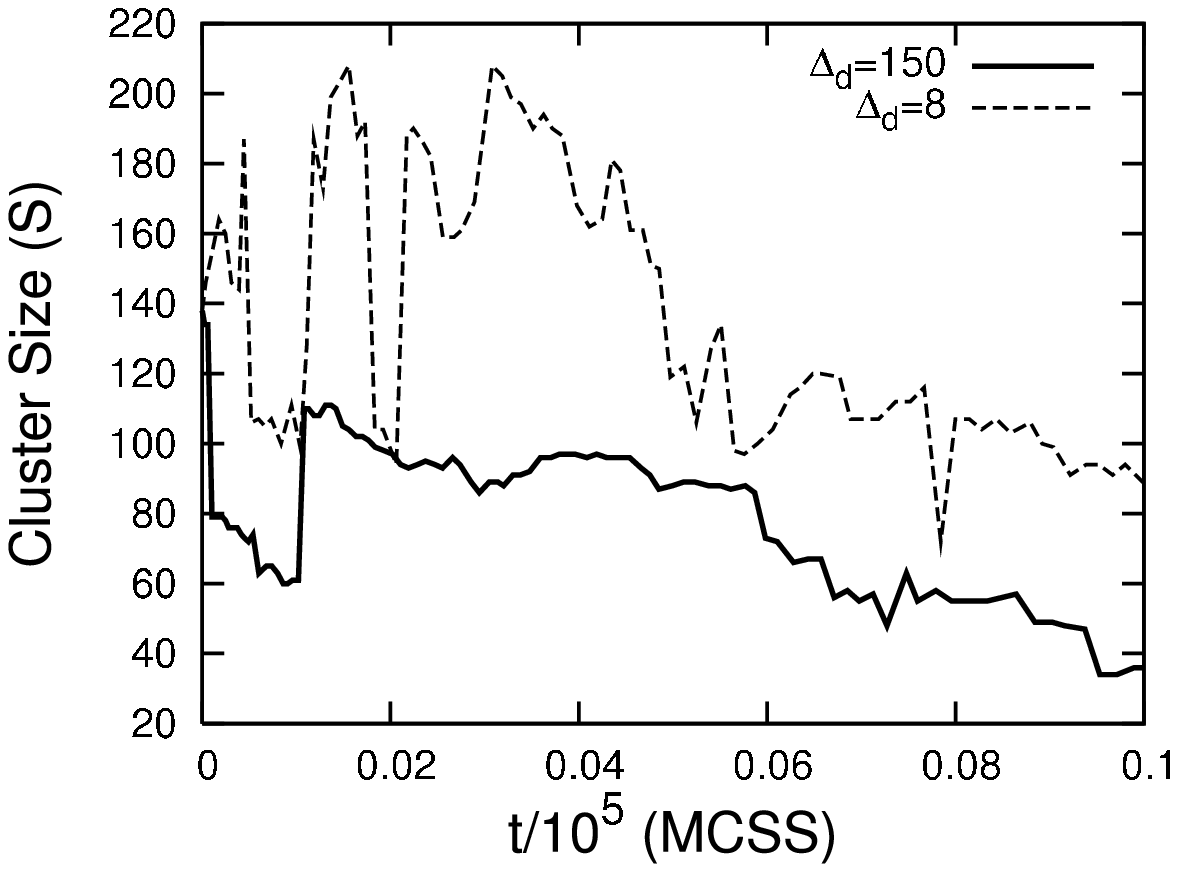}} \\
\end{array}
$$
\end{center}
\caption{The sizes of four clusters (a), (b), (c), and (d) as 
functions of time with and without diffusion. The
solid line is for simulations without diffusion. In (a) we see 
a cluster shrinking.
The diffusion accelerates the process  for late times.
In (b), (c), and (d) we see combinations of split, coalescence, growth, and
shrinkage. In (b),  a cluster splits, with
diffusion accelerating the split. In (c),  diffusion
creates a series of brief coalescences and splits. 
In (d) diffusion creates a stable coalescence.}  
\label{Shrink1}
\end{figure}

\begin{figure}[!ht]
\begin{center}
$$
\begin{array}{cc}
    {\rm (A) } & {\rm (B)} \\
    {\includegraphics[scale=0.6]{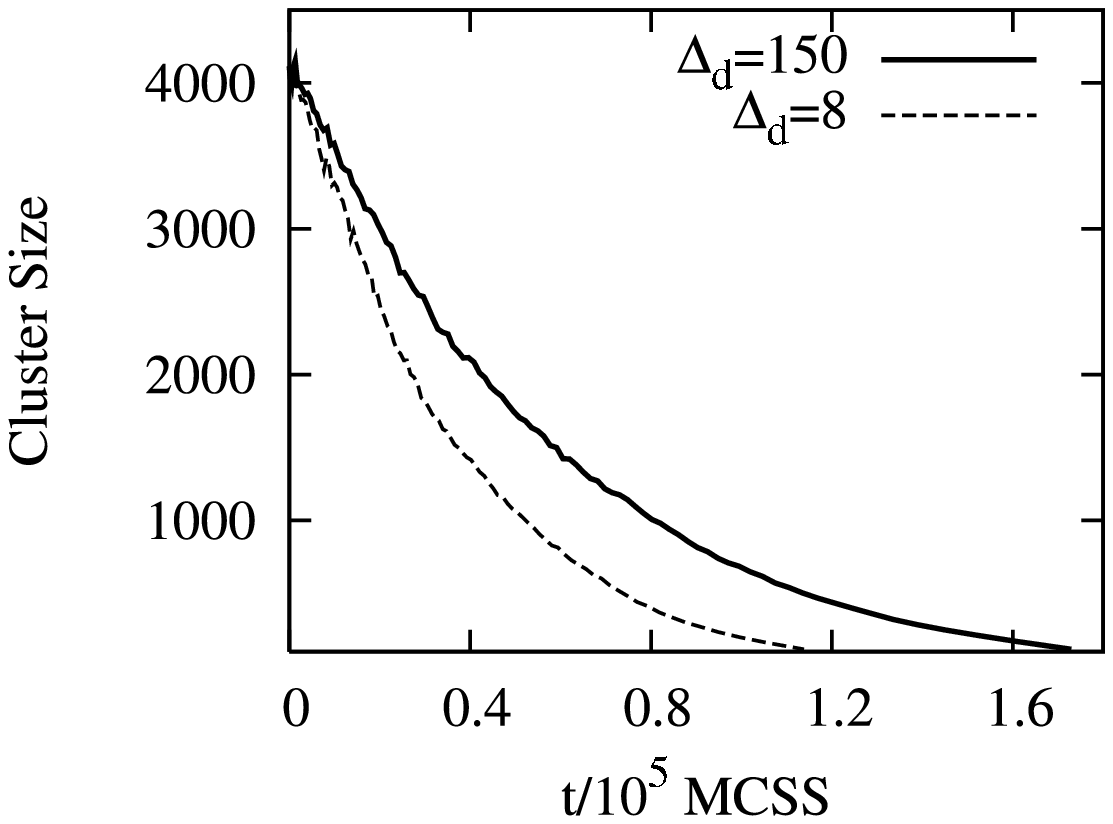}} &
    {\includegraphics[scale=0.6]{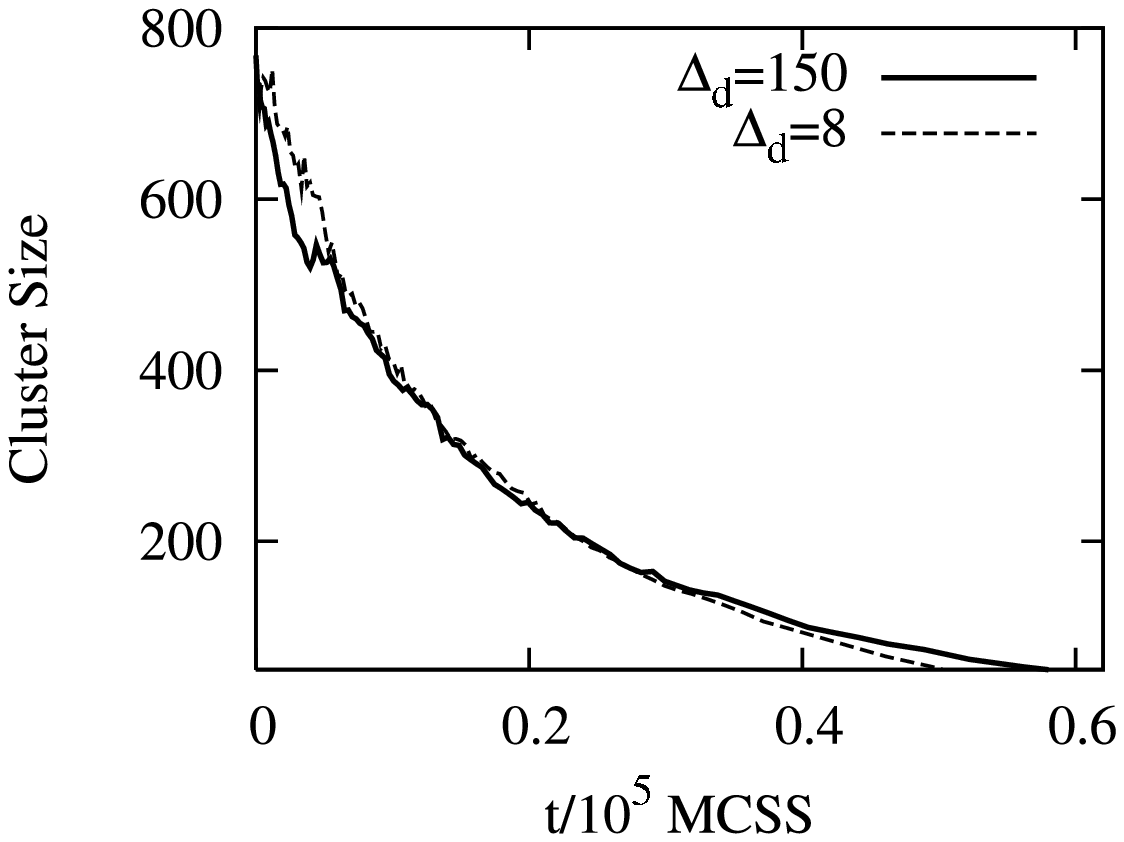}} \\
    {\rm (C) } & {\rm (D)} \\
    {\includegraphics[scale=0.6]{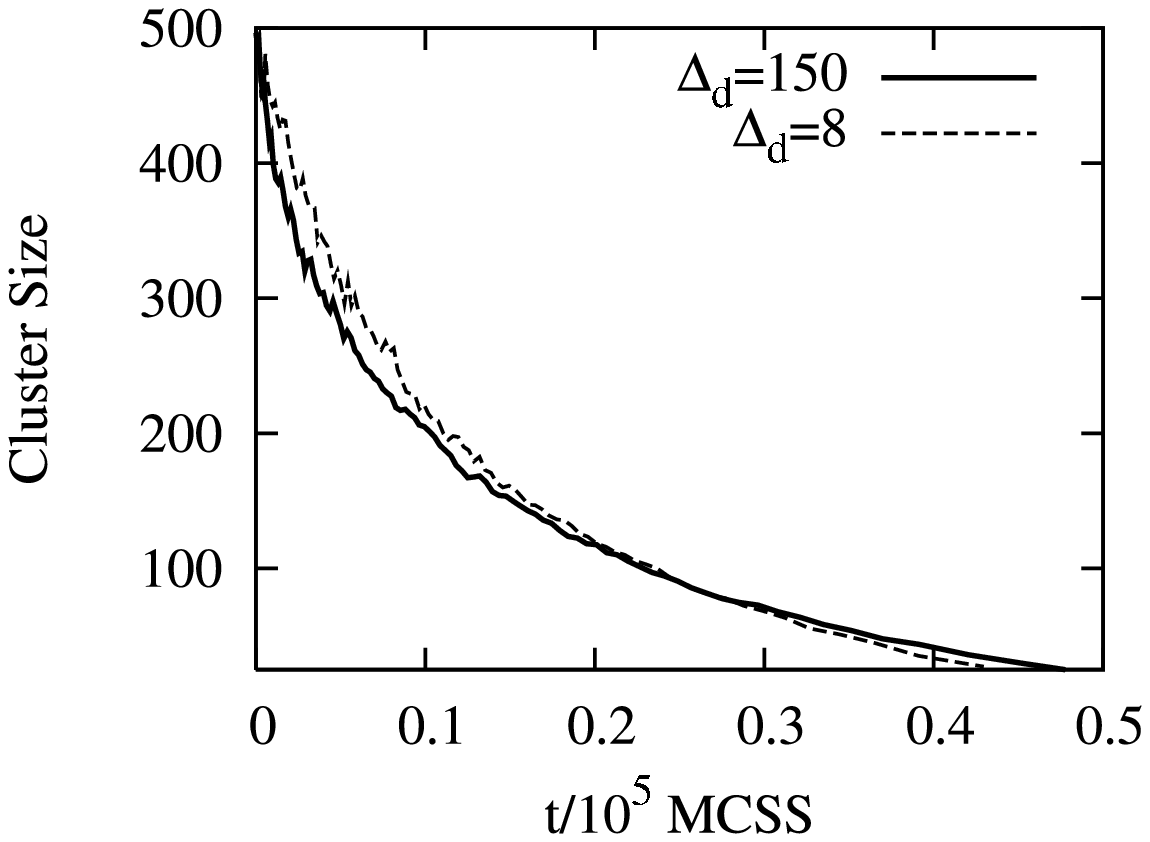}} &
    {\includegraphics[scale=0.6]{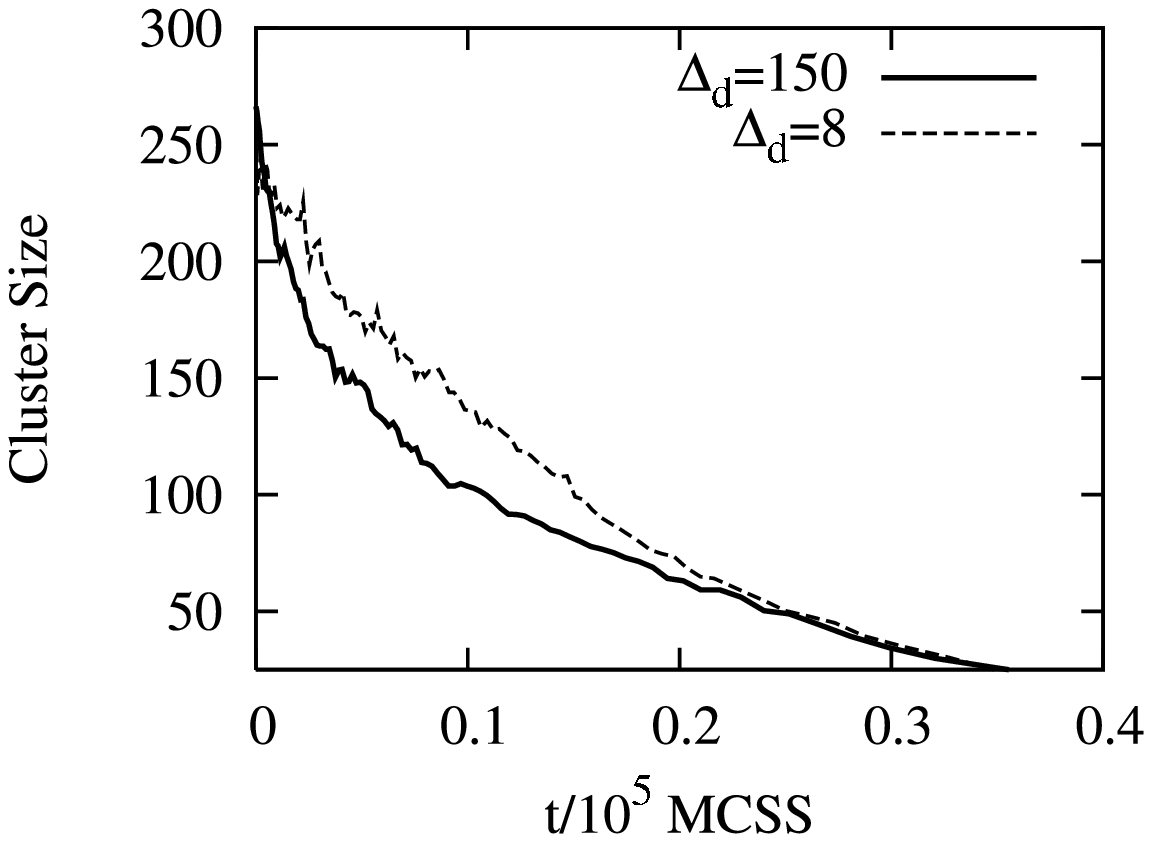}} \\
\end{array}
$$
\end{center}
\caption{Time evolution of the largest cluster,  each picked out from
the simulations with $\xi_{\rm init}$= (A) 4.96, 
(B) 2.35, (C) 1.96, and (D) 1.55, and each averaged
over 100 runs. Solid lines represent simulations
without diffusion, and dashed lines represent simulations with diffusion.}
\label{TLargest1}
\end{figure}

Observation of the dynamics of individual clusters (Figs.~\ref{Shrink1} 
and \ref{TLargest1}) shows that individual clusters 
picked out for observation have similar dynamics as the 
overall system they belong to.
The effect of diffusion on the dynamics of those clusters varies with 
the details of the initial configurations. 
In other words, the effect of diffusion on individual clusters depends
on the {\it size distributions}, i.e., on the total
environment surrounding each cluster.

We believe our observations can help interpret the details of 
experiments such as those reported 
in Refs.~\onlinecite{BAR95},~\onlinecite{BAR96}, and~\onlinecite{HEY01}.

\begin{acknowledgments}
\noindent The authors would like to thank Ibrahim Abou Hamad,
Gregory Brown, and  Gloria M.\ Buend{\'\i}a 
for useful and insightful discussions and comments on the manuscript, 
and three anonymous Referees for helpful suggestions.
 
This work was supported in part by U.S.\
National Science Foundation grant No. DMR-1104829
and by the State of Florida through the Florida State University
Center for Materials Research and Technology (MARTECH).
\end{acknowledgments}


\begin{thebibliography}{10}
\bibitem{ISING}
The lattice-gas Hamiltonian, Eq.~(\protect\ref{hamil}), 
is linked to the Ising model Hamiltonian,
${\mathcal H}_I=-J\sum_{\langle i,j \rangle}s_is_j-H\sum_{i}s_i$
with $s_i=\pm 1 $ by the transformations,
$c_i(t)=({1}/{2})\left[1+s_i(t)\right]$,
$\phi=4J$, and $\mu=2H+\mu_0$.~\protect\cite{FRA06}
\bibitem{Norms}
The binned densities shown in 
Figs.~\protect\ref{Initconf} and \protect\ref{32HIS1} are normalized by 
summation over the bins $j$ such that 
$\rho_\delta = \sum_s n_s = \sum_j n_j \Delta_j$ and 
$\theta = \sum_s n_s s = \sum_j n_j s(j) \Delta_j$,
where $j=1$ corresponds to $s=1$, $\Delta_j = 2^{j-1}$, and $s(j)$ is an 
average of $s$ over bin $j$.
\end{thebibliography}

\newpage

\begin{table}[!h]
\caption{Parameters for the four simulations. Mean cluster sizes $S$,
 correlation lengths $\xi$, and cluster densities 
$\rho_{\delta}$ are averages over 100 independent runs at coverage
$\theta_{\rm init}=0.35$.}
\label{TAB01}
\begin{center}
\begin{tabular}{ccccc}
\hline 
\hline 
Init. Conf. & $\mu-\mu_0$ (Ads) & $S$ & $\xi$ & $\rho_{\delta}/10^{-3}$\\
\hline 
\hline 
A	 &  0.40 	 & 2169.97 	 &    4.96 	 & 12.47  \\ 
B	 &  1.60 	 &  240.54 	 &    2.35 	 & 21.23  \\ 
C	 &  2.56 	 &  128.92 	 &    1.96 	 & 27.24  \\ 
D	 &  9.76 	 &   60.41 	 &    1.55 	 & 42.36  \\ 
\hline
\end{tabular}
\end{center}
\end{table}

\begingroup
\squeezetable
\begin{table}
\begin{center}
\caption{Correlation length difference $\Delta\xi$,
mean cluster size fraction $\delta S$, 
 and time difference $\Delta t$. Primes signify 
simulations with diffusion.}
\label{TAB03}
\begin{tabular}{cccccc}
\hline
\hline
Run&~~~$\xi_{\rm init}$~~~  & ~~~$\theta$~~~  &~~~$\Delta \xi=\xi'-\xi$~~~   &~~~$\delta S=(S'-S)/S$~~~ &$~~~\Delta t=(t-t')/10^5$~~~ \\
\hline
\hline
A&  4.96    &0.35 &0.000 &0.000 &0.000  \\
 &      &  0.32 &      0.069 &      0.011 &     0.016	\\
 &	&  0.28 &      0.097 &      0.035 &     0.038	\\
 &	&  0.25 &      0.083 &      0.055 &     0.066	\\
 & 	&  0.18 &      0.101 &     $-$0.054 &     0.146	\\
 &	&  0.12 &      0.043 &      0.052 &     0.237	\\
 &	&  0.05 &      0.004 &     $-$0.112 &     0.592	\\
\hline
B& 2.35	&0.35 &0.000  &0.000  &0.000  \\
 &      &  0.32 &      0.190 &      0.235 &    $-$0.001	\\
 &	&  0.28 &      0.216 &      0.194 &     0.000	\\
 &	&  0.25 &      0.220 &      0.044 &     0.001	\\
 &	&  0.18 &      0.194 &      0.075 &     0.009	\\
 &	&  0.12 &      0.143 &      0.161 &     0.021	\\
 &	&  0.05 &    $-$0.003 &      0.247 &     0.117	\\
\hline
C& 1.96	&0.35 &0.000  &0.000  &0.000  \\
 &       &  0.32 &      0.275 &      0.268 &    $-$0.003	\\
 &	&  0.28 &      0.304 &      0.202 &    $-$0.004	\\
 &	&  0.25 &      0.300 &      0.173 &    $-$0.004	\\
 &	&  0.18 &      0.253 &      0.134 &    $-$0.002	\\
 &	&  0.12 &      0.167 &      0.108 &     0.003	\\
 &	&  0.05 &     $-$0.000 &      0.072 &     0.069	\\
\hline
D& 1.55	&0.35 &0.000  &0.000  &0.000  \\
 &       &  0.32 &      0.421 &      0.392 &    $-$0.005	\\
 &	&  0.28 &      0.447 &      0.336 &    $-$0.008	\\
 &	&  0.25 &      0.431 &      0.275 &    $-$0.012	\\
 &	&  0.18 &      0.339 &      0.257 &    $-$0.017	\\
 &	&  0.12 &      0.225 &      0.236 &    $-$0.018	\\
 &	&  0.05 &     $-$0.009 &      0.099 &     0.030	\\
\hline
\end{tabular}
\end{center}
\end{table}
\endgroup

\newpage

\bfig
{\includegraphics[scale=0.6]{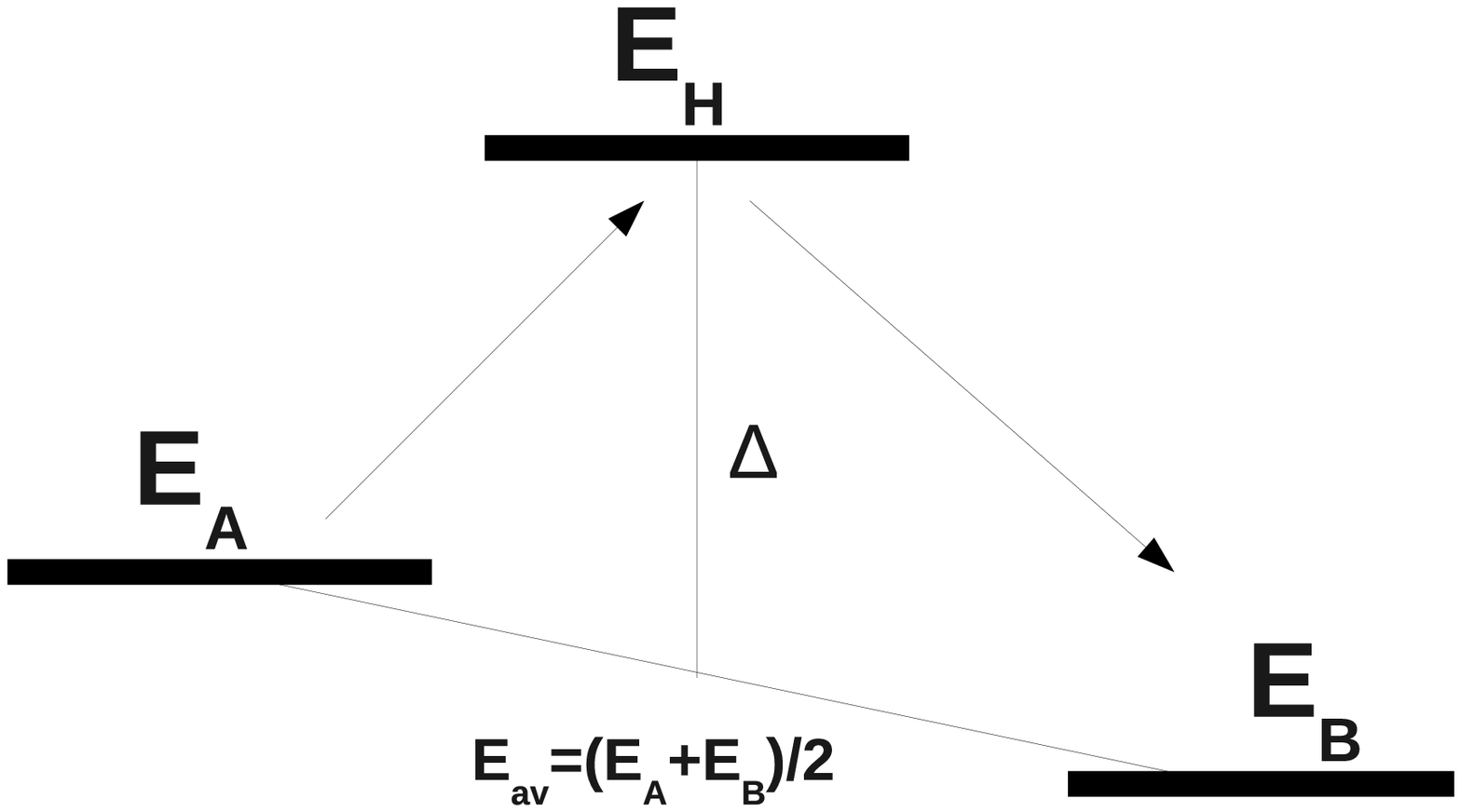}}
\caption{Energy barrier for transitions between states A and B. $E_{\rm H}$ 
is the maximum energy (saddle-point energy) along the transition path.}
\label{picbarr}
\efig

\end{document}